\documentclass{INTERSPEECH2023}
\usepackage{graphicx}
\usepackage{amsmath,amsfonts,bm}
\usepackage{xcolor}
\usepackage{multirow}
\usepackage{booktabs}
\usepackage{url}
\usepackage{algorithm,algpseudocode}

\interspeechcameraready

\title{Edit Distance based RL for RNNT decoding}
\name{Dongseong Hwang, Changwan Ryu, Khe Chai Sim}
\address{Google, U.S.A}
\email{\{dongseong, changwan, khechai\}@google.com}

\begin{document}

\maketitle

\def\x{{\mathbf x}}
\def\y{{\mathbf y}}
\def\L{{\cal L}}

\newcommand{\RDistBayes}{\hat{R}_*}
\newcommand{\REmp}{\hat{R}}

\newcommand{\gL}{\mathcal{L}}
\newcommand{\vx}{\bm{x}}
\newcommand{\vy}{\bm{y}}
\newcommand{\lm}{lm_u}
\newcommand{\am}{am_t}

\begin{abstract}
RNN-T is currently considered the industry standard in ASR due to its exceptional WERs in various benchmark tests and its ability to support seamless streaming and longform transcription. However, its biggest drawback lies in the significant discrepancy between its training and inference objectives. During training, RNN-T maximizes all alignment probabilities by teacher forcing, while during inference, it uses beam search which may not necessarily find the maximum probable alignment. Additionally, RNN-T's inability to experience mistakes during teacher forcing training makes it more problematic when a mistake occurs in inference. To address this issue, this paper proposes a Reinforcement Learning method that minimizes the gap between training and inference time. Our Edit Distance based RL (EDRL) approach computes rewards based on the edit distance, and trains the network at every action level. The proposed approach yielded SoTA WERs on LibriSpeech for the 600M Conformer RNN-T model.
\end{abstract}
\section{Introduction}
The Recurrent Neural Network-Transducer (RNN-T)~\cite{Graves2012} has demonstrated significant success in both academic and industrial settings for automatic speech recognition (ASR)~\cite{sainath2020streaming,li2021better,narayanan2019recognizing,hwang2022pseudo}. This can be attributed to several factors, including the availability of multiple public benchmarks that have established the RNN-T model as state-of-the-art (SoTA) for ASR~\cite{zhang2020pushing,xu2021self,chung2021w2v,zhang2022bigssl}, such as LibriSpeech~\cite{panayotov2015librispeech}, SpeechStew~\cite{chan2021speechstew}, and Multi-lingual LibriSpeech~\cite{pratap2020mls}. Additionally, the RNN-T model offers seamless support for streaming ASR~\cite{chiu2018state} and longform utterances~\cite{lu2021input}, which are crucial requirements for many real-world applications. In comparison, other popular models such as the Connectionist temporal classification (CTC)~\cite{graves2006connectionist} and attention-based models~\cite{chan2016listen} have limitations. Specifically, CTC models generally exhibit higher word error rates (WER) than RNN-T models~\cite{chung2021w2v}, while attention-based models are challenging to support streaming requirements and require non-trivial modifications to support longform inputs~\cite{radford2022robust}.

Despite its advantages in streaming and longform support, the RNN-T model is not without its limitations. Unlike attention-based ASR models, the RNN-T model requires teacher-forcing during training, which means that only ground truth labels are used regardless of the predicted output. As a result, the RNN-T model lacks experience in recovering from errors during training. This is in contrast to the scheduled sampling approach~\cite{bengio2015scheduled,chan2016listen} used by attention-based models, where ground truth tokens are replaced by predicted output tokens randomly. However, the RNN-T model's training objective is to maximize the probabilities of all possible alignments, which is incompatible with scheduled sampling. The CTC model exhibits similar limitations to the RNN-T model.

An additional limitation of the RNN-T model is the significant discrepancy between its training objective and inference method. While beam search is the standard approach for ASR inference, there is no guarantee that it will identify the highest log likelihood hypothesis. The RNN-T model's training objective is to maximize the probability of all possible alignments, which differs substantially from the beam search approach. In contrast, the cross-entropy (CE) loss utilized by attention-based ASR models maximizes the token-level probability, which aligns more closely with the beam search algorithm.

Minimum Word Error Rate Training (MWER) training was proposed as a solution to the problems caused by exposure bias due to teacher-forcing and the discrepancies between the training objective and inference method~\cite{prabhavalkar2018minimum}. MWER is commonly employed in industry as a crucial step in production models, typically following RNN-T pretraining. Nevertheless, this training approach is a sentence-level policy gradient~\cite{williams1992simple}, which results in poor sample efficiency.

We introduce a novel reinforcement learning (RL) technique for RNN-T decoding. Our research offers two noteworthy contributions:

\begin{enumerate}
\item We present a functional RL algorithm for the RNN-T model that generates a training signal for each token rather than for the entire sentence. Our proposed approach leads to notable improvements in the SoTA performance on LibriSpeech WERs.
\item We propose a novel reward engineering technique that utilizes edit distance to proficiently train both emission and blank actions within RNN-T models, thereby achieving direct minimization of the Word Error Rate (WER) metric.
\end{enumerate}

\section{Related work}

The automatic speech recognition (ASR) problem is widely recognized to exhibit exposure bias as a result of teacher forcing and the objective gap between training and inference. Notable adaptations of reinforcement learning (RL) have been proposed for ASR, including the Minimum Word Error Rate (MWER) training~\cite{prabhavalkar2018minimum} approach, which is an ASR version of the REINFORCE algorithm~\cite{williams1992simple}, the most basic RL algorithm introduced in 1992. MWER works by generating the top k hypotheses using beam search and then providing a reward to hypotheses with a better than average WER and a penalty to those with a worse than average WER. MWER has training signal at the sentence level, meaning that all tokens in the same hypothesis are rewarded or penalized together. As WER of SoTA RNN-T models is already less than $10\%$, many of the tokens in the hypotheses are correct but receive an unfair penalty, which can hinder model training. For this reason, both the REINFORCE algorithm and MWER are known to suffer from poor sample efficiency.

In 1999, actor-critic algorithms~\cite{konda1999actor} were introduced as a solution to the issue of poor sample efficiency. This algorithm estimates the value of every action sequence and compute the policy gradient in equation (\ref{eq:rl_grad}) for every action. This approach provides a training signal for each action, resulting in improved sample efficiency. However, a new challenge arises: how to estimate the value. This question has been a major focus of research in RL since the introduction of actor-critic algorithms.

The Optimal Completion Distillation (OCD) method~\cite{sabour2018optimal} provides an innovative solution to the value estimation problem. By leveraging the ability to compute the edit distance for every token, OCD is able to calculate the exact Q-value~\cite{watkins1992q} instead of relying on Q-value estimation. Although OCD works well for character tokens, it is not feasible to apply this method to subword units due to the high computational cost of assigning a Q-value to each subword unit per action. Additionally, if the hypothesis predicts an incorrect subword unit, computing the Q-value for all subword units on the incorrect action becomes a non-trivial problem.

There was a previous attempt to adapt a gradient policy with token-level reward to attention-based ASR in~\cite{tjandra2019end}. Our work builds upon this previous study but focuses on RNN-T~\cite{Graves2012} model. While the earlier paper assigned negative rewards to only the last token, we identified this approach as problematic. Instead, we propose assigning negative rewards to every incorrect action and also propose a value assignment method for blank tokens.
\section{Methods}

\subsection{Training}

Reinforcement learning (RL) generates data via a behavior policy and subsequently computes policy gradient through reward and target policy~\cite{williams1992simple, konda1999actor}. In our study, the target policy is a Recurrent Neural Network-Transducer (RNN-T)~\cite{Graves2012} model that generates a softmax distribution of subword IDs (i.e., WordPiece~\cite{wu2016google}), while the behavior policy is a beam search algorithm based on the RNN-T model. We configure the behavior policy's beam search algorithm to emulate ASR inference time, as our training objective is to optimize the ASR inference process itself.

The training procedure for our model consists of two stages. Firstly, we conduct regular training of RNN-T model. Once a converged checkpoint is obtained, we proceed to further fine-tune the model using the RL objective.

\begin{figure}[t]
  \centering
  \includegraphics[width=\linewidth]{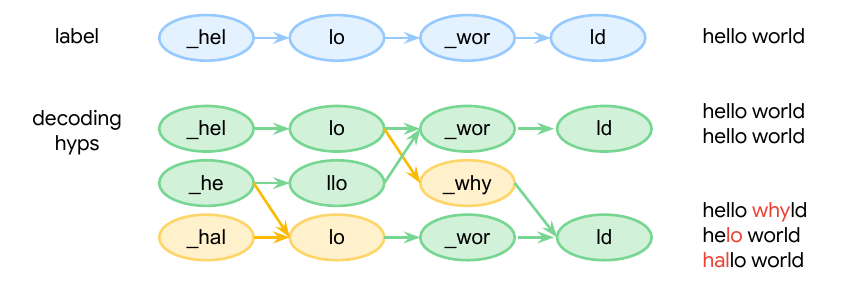}
  \caption{The schematic diagram illustrates how beam search generates hypotheses. Incorrect subwords are highlighted in red and assigned negative rewards, while correct ones receive positive rewards.}
  \label{fig:beam_search}
\end{figure}

\begin{figure}[t]
  \centering
  \includegraphics[width=\linewidth]{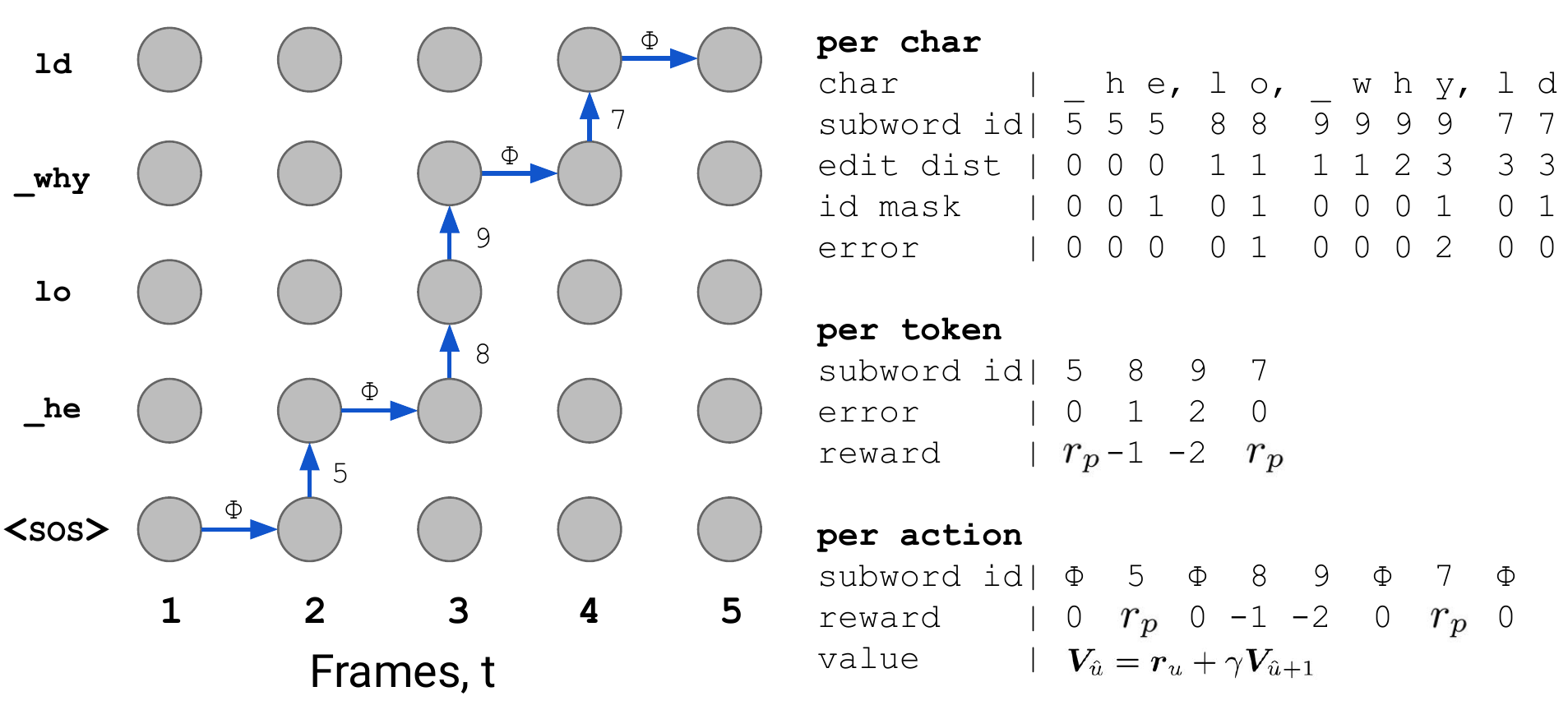}
  \caption{This schematic diagram depicts the process of computing the value for the RL loss, based on the hypothesis 'helo whyld' generated by beam search decoding.}
  \label{fig:value}
\end{figure}

In the context of our study, the process of beam search involves the use of labels, $\y_U$, and audio features, $\x_T$, to generate a hypothesis, $\tilde{\y}_U$, and an associated action sequence, $\hat{\y}_{\hat{U}}$, where $(\y, \phi) \in \hat{\y}$ and $\hat{u}$ represents the index of the action sequence. This process is illustrated in Figure \ref{fig:value}.

Once the hypothesis, $\tilde{\y}_U$, is obtained, we can compute the reward, $\bm{r}_U$, using the edit distance metric. Specifically, we define the error, $e_u$, as the increase in edit distance resulting from a given action. If an action does not contribute to an increase in edit distance, we assign a positive value, $r_p$, as the reward. The specific value of the hyper parameter $r_p$ can vary, and we found that $r_p = 0.1$ performed effectively in the experimental results presented in Section \ref{ssec:abl}.

\begin{equation}
\bm{r}_u=
\begin{cases}
  -e_u, & \text{if } e_u > 0 \\
  r_p, & \text{otherwise} \label{eq:reward}
\end{cases}
\end{equation}

In Figure \ref{fig:value}, we present a TPU-friendly approach to computing rewards, which involves a scatter operation to distribute subword IDs into their constituent characters, followed by computation of the edit distance and error at the character level. The errors are then gathered into token-level values, which are used to compute the reward.

Once the reward, $\bm{r}_u$, is obtained, we can compute the corresponding value, $\bm{V}_{\hat{u}}$, using a discount factor, $\gamma$. The reward, $\bm{r}_u$, is assigned only to emission actions (i.e., $y_k \in \y$), whereas the value, $\bm{V}_{\hat{u}}$, is assigned to both emission and blank actions (i.e., $\hat{y_k} \in \hat{\y} = (\y, \phi)$), which together constitute the action space. It is noteworthy that the value computed based on edit distance offers training signals for both emission and blank actions. It naturally incentivizes or disincentivizes blank actions associated with a good or bad emission action, respectively. In our experiments presented in Section \ref{ssec:abl}, we found that setting $\gamma = 0.95$ provided effective results. 

\begin{align}
\displaystyle \bm{V}_{\hat{u}} &= \bm{r}_{u} + \gamma \bm{V}_{\hat{u}+1} \label{eq:value}
\end{align}

Using the action sequence, $\hat{\y}_{\hat{U}}$, and the corresponding value, $\bm{V}_{\hat{u}}$, we can compute the policy gradient~\cite{Schulmanetal_ICLR2016}. In equation (\ref{eq:rl_grad}), $P_\theta (\hat\y_{\hat{u}}|\hat\y_{\hat{u}-1}, \x_T)$ refers to the probability of subword ID by the RNN-T model.

\begin{align}
\nabla_\theta J(P_\theta) = \mathbb{E}_{\tau \sim P_\theta} [\sum^{\hat{U}}_{\hat{u}=0} \nabla_\theta \log P_\theta (\hat\y_{\hat{u}}|\hat\y_{\hat{u}-1}, \x_T) V_{\hat{u}}]
\label{eq:rl_expected_grad}
\end{align}

An approximation for the expectation can be obtained by calculating the sample mean. We gather a collection of trajectories denoted as $\mathcal{D} = \{\tau_i\}_{i=1,...,N}$ where each trajectory is obtained by beam search using the policy $P_\theta$. The estimation of the policy gradient can be achieved using equation (\ref{eq:rl_grad}), where the number of trajectories in $\mathcal{D}$ (denoted as $|\mathcal{D}|$ or $N$) corresponds to the number of top k hypotheses obtained through beam search.

\begin{align}
\nabla_\theta J(P_\theta) \approx \frac{1}{|\mathcal{D}|} \sum_{\tau \in \mathcal{D}} \sum^{\hat{U}}_{\hat{u}=0} \nabla_\theta \log P_\theta (\hat\y_{\hat{u}}|\hat\y_{\hat{u}-1}, \x_T) V_{\hat{u}}
\label{eq:rl_grad}
\end{align}

The high variance of the gradient estimator is a significant challenge for policy gradient methods~\cite{wu2018variance}. This issue stems, in part, from the challenge of assigning value to the actions that influenced future rewards. This difficulty mainly arises due to the high level of noise in the estimated value, making it challenging to determine whether a given action is beneficial or detrimental. Notably, this problem has received significant attention in the field of RL~\cite{Schulmanetal_ICLR2016,van2016deep,wu2018variance}. In contrast, our proposed method does not have this challenge by computing the value based on edit distance, which enables us to determine precisely whether a given action is advantageous or disadvantageous. Our reward proposal assigns negative values to incorrect actions, which is a relatively uncommon in RL situations.

Since $\bm{V}_{\hat{u}}$ is a value that is detached from the neural network via stop gradient in equation (\ref{eq:rl_grad}), we can compute the RL loss using the following formulation:
\begin{align}
\displaystyle \gL_\text{RL} &= - J(P_\theta) = \frac{1}{N} \sum_{\tau \in \mathcal{D}} \sum^{\hat{U}}_{\hat{u}=0} - \log P_\theta (\hat\y_{\hat{u}}|\hat\y_{\hat{u}-1}, \x_T) V_{\hat{u}} \label{eq:rl_loss}
\end{align}

In order to promote stability during training, we incorporate the RL loss with the RNN-T loss. In our study, we found that setting the weighting factor $\lambda = 0.5$ provided effective results in Section \ref{ssec:abl}.
\begin{align}
\displaystyle \gL_\text{total} &= \lambda \gL_\text{RL} + \gL_\text{RNN-T} \label{eq:tot_loss}
\end{align}

Algorithm \ref{alg:edrl} outlines all the necessary steps for computing the RL loss.
\begin{algorithm}[htb]
\caption{Calculate the Edit distance RL loss}\label{alg:edrl}
\begin{algorithmic}
\Require input audio features, $\x_T$ and labels, $\y_U$.
\State 1. Perform beam search to decode the top k hypotheses, as shown in Figure \ref{fig:beam_search}.
\State 2. Compute the reward per token based on the edit distance using equation (\ref{eq:reward}).
\State 3. Compute the value per action using equation (\ref{eq:value}) as shown in Figure \ref{fig:value}.
\State 4. Calculate the RL loss, $\gL_\text{RL}$, using equation (\ref{eq:rl_loss}).
\end{algorithmic}
\end{algorithm}

\subsection{Inference}

Beam search is a commonly used technique for Automatic Speech Recognition (ASR) models. In our approach, we directly optimize the behavior policy during inference, which necessitates the use of beam search. Consequently, we use beam search during the RNN-T decoding process. Our beam search configuration is set with a beam expansion size of 5 and a top-k value of 4.

\section{Experiments}
\label{sec:exp}

\subsection{Experiment Setup}
\label{sec:experiment_setup}

The Lingvo~\cite{shen2019lingvo} open-source machine learning framework was employed for all experiments conducted in this study.

\subsubsection{Model Architecture}
\label{sec:model_architecture}
In this study, we employed the W2v-BERT Conformer XL model ($0.6$B) introduced in~\cite{zhang2020pushing}. The audio input feature is defined as log mel, comprising 80 dimensions, with a stride of $10ms$ and a window of $25ms$. For the labels, $1k$ WordPiece subword units~\cite{wu2016google} were utilized. The RNN-T model functions as a policy network, with its actions being the subword IDs.

The audio encoder is composed of $24$ Conformer blocks~\cite{conf} with a model dimension of $1024$. The self-attention layer is made up of $8$ heads with $128$ left and right context length, and the convolution kernel size is $5$. The decoder consists of a 2-layer LSTM label encoder with $2048$ units projected down to $640$ output units, and a joint network with a single feed-forward layer with $640$ units. The model has a total of $0.6$B weights.

\subsubsection{Data}
\label{ssec:data}

The LibriSpeech dataset~\cite{panayotov2015librispeech}, which comprises 960 hours of audio and corresponding human transcriptions, is used for all of our experiments such as both RNN-T training and RL training.

\subsubsection{Training procedure}
\label{ssec:train}

In this study, the baseline model underwent a pre-training phase as described in the W2v-BERT paper~\cite{zhang2020pushing}, consisting of a two-step process: firstly, W2v-BERT pretraining using the Libri-Light dataset~\cite{kahn2020libri} (60,000 hours), and secondly, RNN-T finetuning using the LibriSpeech dataset~\cite{panayotov2015librispeech} (960 hours). The model was trained using a batch size of 256 and was found to converge at 15k steps. Once the model has converged, we then further finetune it using the RL loss defined in equation (\ref{eq:tot_loss}). The RL finetune was found to converge at an additional 7k steps.  The RNN-T pretraining was conducted on 16 TPU V3 cores~\cite{jouppi2017datacenter} for a duration of half day, followed by RL finetuning on 32 TPU V3 cores for an additional half day.

\subsection{Experiment results}
\label{sec:results}

We first train the RNN-T baseline model (\texttt{B0}), and then perform further finetuning using the RL objective (\texttt{E1}). For comparison, we also perform finetuning using the MWER~\cite{prabhavalkar2018minimum} objective (\texttt{E2}). The EDRL method utilizes an action-level policy gradient, while MWER utilizes a sequence-level policy gradient. The results in Table~\ref{tab:result} show that our RL method (\texttt{E1}) outperforms the RNN-T baseline (\texttt{B0}), whereas our RNN-T baseline already outperforms the WER reported for W2v-BERT~\cite{chung2021w2v} using only the LibriSpeech dataset.

In the W2v-BERT paper~\cite{chung2021w2v}, self-training was further conducted via Noisy Student Training~\cite{park2020improved} on the unlabeled Libri-Light dataset~\cite{kahn2020libri} of 60,000 hours, resulting in the achievement of a SoTA WER. Our RL model attains a WER that is similar to the SoTA WER, without leveraging self-training on extensive unlabeled data.

In contrast, even after an extensive hyperparameter search, MWER~\cite{prabhavalkar2018minimum} was not able to achieve better WERs than the RNN-T baseline in our setup.

To the best of our knowledge, this is the first RL algorithm that achieves SoTA performance on the LibriSpeech dataset using an RNNT model.

\begin{table}[htb]
    \centering
    \caption{The table displays LibriSpeech WERs without language model (LM) fusion, and demonstrates that our proposed EDRL method outperforms the RNN-T baseline.} 
    \label{tab:result}
    \begin{tabular}{lcccc}
        \toprule
        {\multirow{2}*{\textbf{Model}}}& \multicolumn{4}{c}{\textbf{WER}} \\
         & {\textbf{dev}} & {\textbf{dev-other}} & {\textbf{test}} & {\textbf{test-other}} \\
        \midrule
        {B0 (RNN-T pretrain)} & ${1.4}$  & ${2.7}$ & ${1.5}$  & ${2.7}$ \\
        \midrule
        {E1 (EDRL, ours)} & ${1.4}$  & $\textbf{2.6}$ & $\textbf{1.4}$  & $\textbf{2.6}$ \\
        {E2 (MWER~\cite{prabhavalkar2018minimum})} & ${1.5}$  & ${2.8}$ & ${1.5}$  & ${2.9}$ \\
        \midrule
        {W2v-BERT XL~\cite{chung2021w2v}} & ${1.5}$  & ${2.9}$ & ${1.5}$  & ${2.9}$ \\
        {Self-training on 60k~\cite{chung2021w2v}} & ${1.3}$  & ${2.6}$ & ${1.4}$  & ${2.7}$ \\
        \bottomrule
    \end{tabular}
\end{table}

\section{Discussion}
\label{sec:dis}

\subsection{Ablation study}
\label{ssec:abl}

As we propose new algorithm, we conduct the extensive ablation study.

\subsubsection{Negative reward}
\label{sssec:neg}

We put forth a negative reward function denoted by $-e_u$ in equation (\ref{eq:reward}). The function implies that each subword ID receives a distinct negative reward ranging from 1 to the length of the subword. To address concerns surrounding varying rewards, we assigned a constant reward of -1 for incorrect subword IDs. However, this approach led to model divergence as it exploited the reward gap. Specifically, the model began to hallucinate at the end of the utterance since the maximum negative reward was constrained to -1. Thus, rather than utilizing ad-hoc reward engineering, we chose to enable the model to undergo training based on raw edit distance.

\subsubsection{Positive reward}

We conducted a sweep of positive rewards, denoted by $r_p$, across the values (1, 0.9, 0.7, 0.5, 0.3, 0.1, 0.05, 0.01). Our findings indicate that a positive reward value of $r_p = 0.1$ demonstrated optimal performance.

\subsubsection{Discount factor}

We found that the discount factor, alongside the learning rate, is one of the most crucial hyper parameters. We conducted a sweep of discount factor values, denoted by $\gamma$, ranging from 0 to 0.999, with the specific values (0, 0.1, 0.5, 0.9, 0.93, 0.94, 0.95, 0.96, 0.98, 0.99). Our findings suggest that a discount factor value of $\gamma = 0.95$ produced the best performance.

\begin{table}[htb]
    \centering
    \caption{The table presents a comparison of the average LibriSpeech WERs for different discount factors, denoted by $\gamma$.} 
    \label{tab:self-sup}
    \begin{tabular}{lc}
        \toprule
        {\textbf{$\gamma$}} & {\textbf{Avg WER}} \\
        \midrule
        {RNN-T} & ${2.04}$  \\
        \midrule
        {$0.0$} & ${2.03}$  \\
        {$0.1$} & ${2.05}$  \\
        {$0.5$} & ${2.04}$  \\
        {$0.9$} & ${2.04}$  \\
        {$0.93$} & ${2.04}$  \\
        {$0.94$} & ${2.03}$  \\
        {$0.95$} & $\textbf{2.01}$  \\
        {$0.96$} & ${2.04}$  \\
        {$0.98$} & ${2.04}$  \\
        {$0.99$} & ${2.04}$  \\
        \bottomrule
    \end{tabular}
\end{table}

\subsubsection{RL loss weight}

We performed a sweep of RL loss weight values, denoted by $\lambda$ in equation \ref{eq:tot_loss}, ranging from 0.003 to 1, with the specific values (0.003, 0.1, 0.5, 1). Our results suggest that an RL loss weight of $\lambda = 0.5$ produced the best performance.

\subsubsection{RNN-T loss weight}

We performed a sweep of RNN-T loss weight values, while keeping the RL loss weight fixed at 0.5, across the range of (0.0, 0.1, 0.2, 0.5, 1.0). Our results suggest that an RNN-T loss weight of $1.0$ produced the best performance.

If the RNN-T loss weight is set to $0.0$, the WERs increase to $100\%$ due to deletion errors, indicating that the RNN-T model generates too many blank actions. Without the RNN-T loss as guidance, it is challenging for our RL proposal to differentiate between emission and blank actions. Consequently, we treat the RL loss as an auxiliary loss to the RNN-T loss, with the RL loss weight set to $\lambda = 0.5$ and the RNN-T loss weight set to $1.0$.

The approach employed by InstructGPT (i.e., ChatGPT)~\cite{ouyang2022training} for utilizing RL loss as an auxiliary loss is analogous. The corresponding weights assigned to cross-entropy (CE) and reinforcement learning (RL) are 27.8 and 1, respectively. The self-contained RL objective for the sequence-to-sequence model is a topic of potential investigation for future research.

\subsection{Limitation}

The motivation behind this study is to narrow the gap between the WER and the Oracle WER, which represents the best WER among the top-k beam search hypotheses. As shown in Table \ref{tab:oracle}, the Oracle WER is significantly better than the regular WER. Therefore, if we can select the oracle hypothesis from the top-k hypotheses, it would be a significant improvement over the SoTA WERs. The previous ranking approach ~\cite{variani2020neural} involves using audio features, joint features, labels, and hypotheses as input to rank the top-k hypotheses. However, this approach only yields marginal improvements in WERs and is not readily adaptable to streaming and long-form ASR. In this study, we attempt to address this challenge using RL, but our improved results are still far from the Oracle WERs.

\begin{table}[htb]
    \centering
    \caption{The table displays LibriSpeech WERs and Oracle WER after RNN-T training} 
    \label{tab:oracle}
    \begin{tabular}{lcccc}
        \toprule
        {\multirow{2}*{\textbf{Model}}}& \multicolumn{4}{c}{\textbf{WER}} \\
         & {\textbf{dev}} & {\textbf{dev-other}} & {\textbf{test}} & {\textbf{test-other}} \\
        \midrule
        {RNN-T WER} & ${1.4}$  & ${2.7}$ & ${1.5}$  & ${2.7}$ \\
        {Oracle WER} & ${0.55}$  & ${1.3}$ & ${0.56}$ & ${1.4}$ \\
        \bottomrule
    \end{tabular}
\end{table}

\section{Conclusions}

In this paper, we proposed a novel approach for improving the performance of RNN-T models on speech recognition tasks using reinforcement learning (RL). Our approach involves directly optimizing beam search for inference time, which leads to better performance compared to the RNN-T baseline and other existing approaches. Our approach represents the first successful utilization of RL for the LibriSpeech dataset, and it has achieved SoTA WERs. Our findings suggest that RL has the potential to enhance the accuracy of RNN-T models in speech recognition tasks, paving the way for further research in this area.



\bibliographystyle{IEEEtran}
\bibliography{ref}

\end{document}